\documentstyle[12pt]{article}
\begin{document}

\input epsf
\ifx\epsfbox\UnDeFiNeD\message{(NO epsf.tex, FIGURES WILL BE IGNORED)}
\def\figin#1{\vskip2in}
\else\message{(FIGURES WILL BE INCLUDED)}\def\figin#1{#1}\fi
\def\footnotefont{\tenpoint}

\parindent 25pt
\overfullrule=0pt
\tolerance=10000
\def\Re{\rm Re}
\def\Im{\rm Im}
\def\titlestyle#1{\par\begingroup \interlinepenalty=9999
     \fourteenpoint
   \noindent #1\par\endgroup }
\def\tr{{\rm tr}}
\def\Tr{{\rm Tr}}
\def\half{{\textstyle {1 \over 2}}}
\def\calt{{\cal T}}
\def\ie{{\it i.e.}}
\def\np{Nucl. Phys.}
\def\pl{Phys. Lett.}
\def\pr{Phys. Rev.}
\def\prl{Phys. Rev. Lett.}
\def\cmp{Comm. Math. Phys.}
\def\quart{{\textstyle {1 \over 4}}}
\def\RR{${\rm R}\otimes{\rm R}~$}
\def\NSNS{${\rm NS}\otimes{\rm NS}~$}
\def\RNS{${\rm R}\otimes{\rm NS}~$}
\def\calf{${\cal F}$}

\baselineskip=17pt
\pagestyle{empty}
{\hfill DAMTP-97-44}
\vskip 0.1cm
{\hfill hep-th/9705023}
\vskip 0.4cm
\centerline{CONTACT TERMS, SYMMETRIES AND D-INSTANTONS.}
\vskip 1cm
 \centerline{ Michael Gutperle\footnote{M.Gutperle@damtp.cam.ac.uk}}
\vskip 0.3cm
\centerline{DAMTP, Silver Street,}
\centerline{ Cambridge CB3 9EW, UK.}
\vskip 1.4cm
\centerline{ABSTRACT}
\vskip 0.3cm
\noindent The scattering of  \NSNS\ antisymmetric tensor states in
the presence of D-instantons in type IIB superstring theory is
studied. It is shown that in order to preserve  gauge invariance, spacetime
supersymmetry and picture changing symmetry the inclusion of  boundary
contact terms for  closed string antisymmetric tensor vertex
operators is necessary. 
\vfill\eject
\pagestyle{plain}
\setcounter{page}{1}

\section{Introduction}

The gauge invariance of  string scattering amplitudes manifests
itself   by
the fact that vertex operators corresponding to longitudinal states
are total derivatives on the world sheet and decouple. Such arguments
can fail either when there is a contribution from the boundary
of moduli space or when the world sheet itself has
boundaries. One important example is given by the Cremmer-Scherk
mechanism \cite{Cremmerscherk} where the \NSNS\ antisymmetric tensor $B_{\mu\nu}$ and the open string
vector state $A_\mu$ mix and only the  combination
$B_{\mu\nu}-F_{\mu\nu}$  is gauge
invariant. Contact terms can be included to justify
analytic continuation in external momenta (the so called `cancelled
propagator' 
argument). On the other hand unphysical states may  fail to decouple in string
scattering amplitudes  and
contact terms are needed to restore gauge invariance and unitarity \cite{greenseiberg}. In
some circumstances kinematic restrictions  make analytic continuation impossible
and contact terms are essential \cite{greenseiberg,seibergb}.
In this paper the scattering of closed string states is analyzed in the presence of
D-instantons. In particular when \NSNS\ antisymmetric tensors (AST) are present  it is
shown that the various symmetries of string perturbation theory make
a boundary contact term necessary. In section \ref{BRST} the
BRST cohomology of the open string sector for D-instantons is
analyzed. In section \ref{gauge} it is  shown  that a \NSNS\ AST tensor gauge transformation
fails to decouple and a boundary term must be added in order to cancel
this contribution. In section \ref{susy} it is shown that a space time
supersymmetry transformation also induces a boundary term and is
cancelled by the same contact term. In section \ref{pic} the
invariance under picture changing of a
closed string scattering amplitude in the one D-instanton sector is
investigated and again the same contact terms are necessary for the
invariance. The amplitude with  two  \NSNS antisymmetric tensors is used as an illustration. In section \ref{conclusions} we
present some conclusions and speculations.
\section{Dirichlet instantons and BRST cohomology}\label{BRST}

The BRST charge $Q_{BRST}$ for the open string is given by 
\begin{equation}
  Q_{BRST}=\oint dz  \left\{c( -{1\over 2} \partial X\partial X
  +{1\over 2}\psi\partial \psi -{1\over 2} \partial \beta \gamma
  -{3\over 2} \beta \partial\gamma)+c\partial c b+ {1\over 2} \gamma
  \psi\partial X -{1\over 4} \gamma^2b\right\},
\end{equation}
where $b,c$ are the anticommuting reparameterization ghosts and
$\beta,\gamma$ are the commuting super-conformal ghosts. 
The space of physical vertex operators is defined as the
cohomology class of BRST closed vertex operators $\{ Q_{BRST},V(z)\}=0$ modulo BRST
exact  vertex operators $V\sim V+\{Q_{BRST},W\}$.
In the
following the $\beta,\gamma$ system is bosonized 
introducing a scalar $\phi$ which background charge $2$ and a pair of
weight $(1,0)$ fermions $\eta,\xi$ \cite{FMS}
\begin{equation}\label{bosonization}
  \beta = e^{-\phi}\partial\xi,\quad \gamma= \eta e^{\phi}
\end{equation}
In terms of the  bosonized fields  the BRST charge is given by 
\begin{equation}\label{BRSTbose}
  Q_{BRST}=\oint dz  \left\{c( -{1\over 2} \partial X\partial X
  +{1\over 2}\psi\partial \psi-{1\over 2} (\partial \phi)^2-2
  \partial^2\phi)+c\partial c b +{1\over 2} e^\phi \eta \psi\partial X
  -{1\over 4} \eta \partial \eta b\right\}.
\end{equation}
Note the fact that  the zero mode of $\xi$ does not enter in the bosonization
formula (\ref{bosonization}), which is the starting point for the picture changing
operation \cite{FMS}. The $\beta,\gamma$ system
is a commuting first order system which has infinitely many distinct
vacua (pictures) which are labelled by the superghost charge $j=\oint \partial \phi$.

The picture changing operator $\cal{X}$ which maps a vertex operator in a picture
$s$ to  a vertex operator in
picture $s+1$, ${\cal{X}}: V_s\to V_{s+1}$ by 
\begin{equation}\label{picchange}
  V_{s+1}(z)=\{ Q_{BRST},\xi V_s(z)\}.
\end{equation}
In the standard Neumann open string theory the (fixed) vertex operator
for the massless vector state
is given by in the $s=-1$ and $s=0$ picture respectively
\begin{equation}\label{Vneumann}
  V_{-1}=ce^{-\phi}\psi^\mu e^{ikX},\quad V_{0}= c(\partial X^\mu +ik_\rho\psi^\rho\psi^\mu)e^{ikX}.
\end{equation}
It is easy to see that the two vertex operators are indeed related
by (\ref{picchange}) and that the vertex operator represent physical
states in an appropriate gauge if
$k_\mu\zeta^\mu=0$ and $k^2=0$. 
In  the Neveu-Schwarz sector there is a simpler and less sophisticated
description of the picture changing. One can define
two pictures called $F_1$ and $F_2$  \cite{gsw}. For open strings the
NS-sector states
$F_2$ picture obeys the mass shell condition  $(L_0-{1\over
  2})\mid \phi\rangle=0$ and $G_r\mid \phi\rangle=L_n\mid
\phi\rangle=0$ for $r>0,n>0$. Because $G_{1/2}G_{-1/2}\mid \phi
\rangle_{F_2}=\mid \phi \rangle_{F_2}$ one can define an open string state in the
$F_{1}$ picture defined by $\mid \phi \rangle_{F_1}=G_{-1/2}\mid \phi
\rangle_{F_2}$. For the open string with Neumann boundary conditions
the massless vector states take the following form in the two
pictures.
\begin{eqnarray}
  \mid \zeta_\mu,k\rangle_{F_2}&=&\zeta_\mu \psi_{-1/2}^\mu\mid
  k\rangle,\label{openF2}\\
\mid \zeta_\mu,k\rangle_{F_1}&=&\zeta_\mu( a_{-1}^\mu+ik_\rho
  \psi_{-1/2}^\rho \psi_{-1/2}^\mu)\mid
  k\rangle.\label{openF1}
\end{eqnarray}
Which shows that the state (\ref{openF1}) in the $F_1$ picture is
given by a
worldsheet supersymmetry transformation of the state (\ref{openF2}) in
the $F_2$ picture.

 D-instanton boundary conditions on the open string coordinate $X^\mu$
 are defined by imposing   $X^{\mu}|_{\sigma=0}=Y_1^\mu$
and $X^{\mu}|_{\sigma=\pi}=Y_2^\mu$. The mode expansion is then given by
\begin{equation}
  X^\mu=Y_1^\mu+{Y_2^\mu-Y_1^\mu\over \pi}\sigma
  +\sqrt{2\alpha^\prime}\sum_n {a_n^\mu\over n} e^{-in\tau}\sin(n\sigma).
\end{equation}
The zero mode of the matter energy momentum tensor for an open string
`stretched' between two D-instantons at $Y_1^\mu$ and $Y_2^\mu$ is
given by
\begin{equation}
  L^X_0={(Y_1-Y_2)^2\over 4\pi \alpha^\prime}+\sum_n:a^\mu_{-n}a_{-n\mu}:\;.
\end{equation}
The usual open string momentum for an open string satisfying Neumann
boundary conditions is replaced by the separation
$(Y_1^\mu-Y_2^\mu)/{\alpha^\prime}$ for the D-instanton. 

For a single D-instanton which is located at $y^\mu$ we have
$Y_1^\mu=Y_2^\mu=y^\mu$ 
This implies that the BRST cohomology for D-instanton open string
states is isomorphic to the cohomology of Neumann open strings at zero
momentum \cite{mbg7}.
The open string states do not carry any momentum and the vertex
operator for the level one vector state is given by
\begin{equation}
  V_{-1}=\zeta_\mu\int dx e^{-\phi}\psi^\mu,\quad V_0= \zeta_\mu\int
  dx \partial_n X^\mu.
\end{equation}
It is easy to see that both vertex operators commute with the BRST
charge (\ref{BRSTbose}) without imposing any constraint on the
wavefunction $\zeta_\mu$ which is interpreted as a collective
coordinate corresponding to the shift in the position of the
D-instanton $y^\mu$ in space
time. Polchinski has shown that upon integration over the instanton
position $y^\mu$ all divergences associated with boundary of moduli
space cancel \cite{polchinc}.
There is another level one state which appears in the vertex $V_0$ (\ref{Vneumann})
  \begin{equation}\label{psipsi}
  V^c_0=  \zeta_{\mu\nu}\oint dx \;\psi^\mu\psi^\nu\; e^{iky}.
  \end{equation}
In the case of the D-instanton this is not  BRST invariant for
nonvanishing $\zeta_{\mu\nu}$ since $\{Q_{BRST}, V^c_0\}=\zeta_{\mu\nu}\oint dx
e^\phi \eta \psi^\mu\partial X^\nu$  and therefore unphysical. As we shall
see in the subsequent section this unphysical state will become
important as a contact term which restores closed string symmetries in
the presence of D-instantons.

\section{Gauge invariance}\label{gauge}
The effect of boundaries with D-instanton boundary conditions is
studied in the simplest case where the worldsheet is the upper half
plane $H=\{z|Im(z)>0\}$. Dirichlet boundary conditions are imposed on
all the coordinates and the boundary conditions on the worldsheet
fermions are determined by worldsheet supersymmetry \cite{polchinskia}.
\begin{equation}\label{bcXPsi}
  \left(\partial
  X^\mu(z)+\bar{\partial}X^\mu(\bar{z})\right)\mid_{Im(z)=0}=0,\quad \left(\psi^\mu(z)+\bar{\psi}^\mu(\bar{z})\right)\mid_{Im(z)=0}=0.
\end{equation}
The boundary is given by the real line, open string vertex
operators are written in terms of leftmoving oscillators only and
correlation functions can be evaluated by the well known doubling
procedure \cite{lechtenfeld}.
The vertex operator for the \NSNS\ state in the $(0,0)$ picture is
given by
\begin{equation}\label{vertexNSNS}
 V(\zeta)= \zeta_{\mu\nu}\left(\partial
  X^\mu+ik_\rho\psi^\rho\psi^\mu\right)\left(\bar{\partial}
  X^\nu+ik_\lambda\bar{\psi}^\lambda\bar{\psi}^\nu\right)e^{ikX}(z,\bar{z}).
\end{equation}
The physical state conditions for the massless tensor states imply
that $\zeta_{\mu\nu}k^\mu=\zeta_{\mu\nu}k^\nu=0$ and $k^2=0$. The
tensor $\zeta_{\mu\nu}$ decomposes into the traceless symmetric part (graviton), dilaton and
antisymmetric part. For the graviton and antisymmetric tensor the
gauge invariance associated manifests itself in the fact that upon
the replacement $\delta \zeta^{(1)}=k_\mu\Lambda_\nu$ and $\delta
\zeta^{(2)}_{\mu\nu}=k_\nu\Lambda_\mu$ the vertex operator
(\ref{vertexNSNS}) becomes a total derivative  which
decouples on a compact worldsheet.
\begin{eqnarray}
  V(\delta \zeta^{(1)})&=&\Lambda_\nu\partial\left\{(\bar{\partial}
  X^\nu+ik_\lambda\bar{\psi}^\lambda\bar{\psi}^\nu)e^{ikX}\right\},\label{total1}\\
 V(\delta \zeta^{(2)})&=&\Lambda_\nu\bar{\partial}\left\{(\partial
  X^\nu+ik_\lambda{\psi}^\lambda{\psi}^\nu)e^{ikX}\right\}.\label{total2}
\end{eqnarray}
A gauge transformation on the graviton wavefunction  corresponds to a linearized
coordinate transformation and is given by  $\delta \zeta^{grav}_{\mu\nu}=\delta
\zeta^{(1)}_{\mu\nu}+\delta
\zeta^{(2)}_{\mu\nu}$. For D-instanton boundary conditions the total
derivatives in (\ref{total1}) and (\ref{total2}) produce a boundary term of the form
\begin{equation}
 \int d^2z V(\delta \zeta^{grav})=\Lambda_\nu \int dx \partial_n X^\nu=
 \Lambda_\nu{\partial \over \partial Y_\nu}.
\end{equation}
The boundary operator is proportional to the momentum conjugate to the
position of the D-instanton and the gauge transformation induces an
infinitesimal shift in the position of the D-instanton $\delta
Y_\nu=\Lambda_\nu$.

For the antisymmetric tensor a gauge transformation is given by $\delta \zeta^{AST}_{\mu\nu}=\delta
\zeta^{(1)}_{\mu\nu}-\delta
\zeta^{(2)}_{\mu\nu}$ and yield the following boundary term
\begin{equation}\label{boundaryAST}
   \int d^2z V(\delta \zeta^{AST})=i\Lambda_{[\nu} k_{\mu]} \int dx\;
   \psi^\mu\psi^\nu e^{iky}.
\end{equation}
In order to preserve gauge invariance a boundary term (\ref{psipsi})
has to be added where the wavefunction $\zeta_{\mu\nu}$ transforms
$\delta\zeta_{\mu\nu}=k_{[\mu}\Lambda_{\nu]}$ under gauge
transformations and therefore cancels  a boundary term  (\ref{boundaryAST}).

The boundary conditions (\ref{bcXPsi}) reflect left and
rightmoving degrees of freedom. This leads to nontrivial OPE between
left and rightmovers for closed string vertex operators coming close
to the boundary which can be
used to deduce the coupling of closed and open strings \cite{cardy}. 
For a \NSNS\ AST vertex operator located at $z=x+iy$ the singular part
of the boundary
induced OPE is given by
\begin{eqnarray}\label{opeclosed}
  lim_{y\to 0}V(\zeta^{AST}_{\mu\nu})&\sim& \zeta^{AST}_{\mu\nu}{1\over
  (y)^{k^2+1}}k^2 \psi^\mu\psi^\nu (x)+o(y^0)\nonumber \\
&=& -\left({\partial\over \partial y} y^{k^2}\right)  \zeta^{AST}_{\mu\nu}\psi^\mu\psi^\nu (x)+o(y^0).
\end{eqnarray}
This situation is very similar to the one discussed in
\cite{greenseiberg}, for the integrated vertex operator the total derivative gives $\int
dy{\partial\over \partial y} y^{k^2}= y^{k^2}\mid_{y=0}$. For $k^2>0$
this term vanishes whereas  for $k^2<0$ it is divergent but 
the on-shell condition for the closed string vertex operator enforces
$k^2=0$. The contact term at the
boundary is therefore necessary to cancel the total derivative and to  restore the gauge
invariance.

Note that the contact term (\ref{psipsi}) can be absorbed
into the the AST vertex operator in order to write it partly in a
manifestly gauge invariant fashion. Using the boundary conditions
(\ref{bcXPsi}) the boundary contact term is  reexpressed as a bulk vertex,
\begin{eqnarray}\label{boundbulk}
  \zeta_{\mu\nu}\int dx\; \psi^\mu\psi^\nu e^{iky}&=&\half\zeta_{\mu\nu} \int d^2z
  \partial\left(\bar{\psi}^\mu\bar{\psi}^\nu e^{ikX}\right)- \half\zeta_{\mu\nu} \int d^2z
 \bar{ \partial}\left({\psi}^\mu{\psi}^\nu e^{ikX}\right)\nonumber\\
&=&i\half \zeta_{\mu\nu}k_\rho \int d^2z \left({\psi}^\mu{\psi}^\nu
  \bar{\partial}X^\rho- \bar{\psi}^\mu\bar{\psi}^\nu\partial
  X^\rho\right)e^{ikX}. 
  \end{eqnarray}
 In this
 form it can be added to the part of the AST vertex (\ref{vertexNSNS})
 which fails to decouple under linearized gauge transformations namely
 the terms 
 \begin{eqnarray}
&& i\zeta_{\mu\nu}k_\rho \int d^2z \left( \partial X^\mu
 \bar{\psi}^\rho\bar{\psi}^\nu-\bar{\partial}X^\mu \psi^\rho\psi^\nu\right)e^{ikX}+ \zeta_{\mu\nu}\int dx\; \psi^\mu\psi^\nu e^{iky} \nonumber\\
&=&\half H_{\mu\nu\rho}\int d^2z  \left( \partial X^\mu
 \bar{\psi}^\rho\bar{\psi}^\nu-\bar{\partial}X^\mu
 \psi^\rho\psi^\nu\right)e^{ikX},
\end{eqnarray}
where (\ref{boundbulk}) was used and $H_{\mu\nu\rho}=i\half
\zeta_{[\mu\nu}k_{\rho]}$ is the manifestly gauge invariant field
strength for  the two form AST potential $\zeta_{\mu\nu}$.
\section{Space time supersymmetry}\label{susy}
In the Neveu-Schwarz-Ramond  formalism the space time supersymmetry generators are
constructed from   spin fields and superghosts \cite{FMS}.  
 The charges in the $s=-1/2$
and $s=1/2$ picture are given by
\begin{equation}\label{susycharges}
  Q^a_{-1/2}=\oint dz e^{-1/2\phi}S^a, \qquad Q^a_{1/2}= \oint dz
  e^{1/2\phi} (\gamma^\mu)^a
_b S^b \partial X_\mu.
\end{equation}
Similarly for the rightmoving supercharges $\bar{Q}^a_s$. Because of
the superghost dependent part in (\ref{susycharges}) the spacetime
supersymmetry charges appear in different pictures. Note that the supersymmetry algebra only closes on shell, i.e. modulo
picture changing operation. The anticommutator $\{Q^a_{-1/2},Q^b_{-1/2}\}=\gamma_\mu^{ab}\oint dz
e^{-\phi}\psi^\mu$ is related by picture changing to $\{ Q^a_{1/2},Q^b_{-1/2}\}=\gamma_\mu^{ab}\oint dz \partial
X^\mu$ where the standard space time momentum operator appears on the
right hand side. The supersymmetry transformation on the vertex operators (\ref{vertexNSNS})
is only closing up to total derivatives which again can yield boundary
terms.
Indeed using standard OPE techniques one can show that  the
anti-commutator of the the supersymmetry
charge in the $s=1/2$ picture and the 
leftmoving part  (\ref{vertexNSNS}) is given by
\begin{eqnarray}\label{calc1}
\{Q^a_{1/2},\zeta_\mu \left(\partial
  X^\mu+ik_\rho\psi^\rho\psi^\mu \right)e^{ikX}(z)\}
&=&\zeta_\mu\oint dw {1\over (z-w)^2}e^{1/2\phi} S^b (w)(\gamma^\mu)^a_b
  e^{ikX}(z)\nonumber\\
&-& 
ik_\nu\zeta_\mu\oint dw {1\over z-w}e^{1/2\phi} S^b (w)
  (\gamma^\nu)^a_b\partial X^\mu e^{ikX}(z)\nonumber\\
&+& ik_\nu ik_\rho\zeta_\mu\oint dw {1\over (z-w)^2}e^{1/2\phi} S^b (w)
  (\gamma^\nu \gamma^{\rho\mu})^a_b e^{ikX}(z)\nonumber\\
&+&ik_\rho\zeta_\mu\oint dw {1\over
  z-w}e^{1/2\phi} S^b\partial X_\nu(w)
  (\gamma^\nu\gamma^{\rho\mu})^a_b e^{ikX}(z)\nonumber.\\
\end{eqnarray}
Note that the third term on the right hand side  of (\ref{calc1}) vanishes because of the the
two momentum factors and the physical state condition on the
polarization vector $\zeta_\mu$.
Rearranging  the gamma matrices using $ (\gamma^\nu
\gamma^{\rho\mu})^a_b=\delta^{\nu\rho}(\gamma^\mu)^a_b-\delta^{\nu\mu}(\gamma^\rho)^a_b+(\gamma^{\rho\mu}\gamma^\nu)^a_b$
the other terms can be combined in the following form
\begin{eqnarray}\label{resultleft}
  \{Q^a_{1/2}, \zeta_\mu\left(\partial
  X^\mu+ik_\rho\psi^\rho\psi^\mu \right)e^{ikX}(z)\}&=&-\partial\left(\zeta_\mu
  e^{\phi/2}S^b (\gamma^\mu)^a_b e^{ikX}\right)\nonumber\\
&-&ik_\rho\zeta_\mu e^{\phi/2}S^b \partial
  X_\nu (\gamma^{\rho\mu}\gamma^\nu)^a_b
  e^{ikX}.
\end{eqnarray}
Similarly  the anticommuator of rightmoving supercharge and the
rightmoving part of the vertex (\ref{vertexNSNS}) is
given by
\begin{eqnarray}\label{resultright}
 \{\bar{Q}^a_{1/2}, \bar{\zeta}_\mu\left(\bar{\partial}
  X^\mu+ik_\rho\bar{\psi}^\rho\bar{\psi}^\mu \right)e^{ikX}(\bar{z})\}&=&-\bar{\partial}\left(\bar{\zeta}_\mu
  e^{\bar{\phi}/2}\bar{S}^b (\gamma^\mu)^a_b e^{ikX}\right)\nonumber\\
&-&i k_\rho\zeta_\mu e^{\bar{\phi}/2}\bar{S^b} \bar{\partial}
  X_\nu(\gamma^{\rho\mu}\gamma^\nu)^a_b
  e^{ikX}.
\end{eqnarray}
The supersymmetry transformation of the vertices produces a total
derivative and a term which corresponds to  the linearized supersymmetry
transformation on the spinor in standard open string supersymmetric
Yang-Mills  $\delta \psi =
\Gamma^{\mu\nu}F_{\mu\nu}\epsilon$. The supersymmetries in the closed
string theory are generated by both left and right moving supercharges. 
The boundary conditions  on the fields $X^\mu,\psi^\mu$ (\ref{bcXPsi}) together with
the consistency of the operator product expansion imply the following
boundary condition\footnote{The factor of i appears because analytic
  continuation to euclidean signature for the D-instanton.} on the spin fields \cite{callanb,lia}.
  \begin{equation}\label{bcSS}
  \left(S^a(z)+i(\gamma^{11})^a_b\bar{S}^b(\bar{z})\right)\mid_{Im(z)=0}=0.
  \end{equation}
Defining the two  combinations of supercharges
\begin{equation}
  Q^{a+}_s=Q^a_s+i(\gamma^{11})^a_b\bar{Q}^b_s,\quad  Q^{a-}_s=Q^a_s-(\gamma^{11})^a_bi\bar{Q}^b_s,
\end{equation}
it is easy to see that the  boundary conditions imply that $Q^{a+}$ vanishes at the boundary
and represents the unbroken supersymmetry. Whereas $Q^{a-}$ does not
vanish and corresponds to the broken supersymmetry. It has the
form of a zero momentum open string fermionic vertex operator and
represents the fermionic zero mode produced by the broken
supersymmetry \cite{greengutperlec,greengutperlee}.

We consider the effect of an unbroken  supersymmetry transformation on the \NSNS\ 
antisymmetric tensor (\ref{vertexNSNS}). Combining (\ref{resultleft})
and (\ref{resultright}) gives
\begin{equation}\label{contact1}
  \{\epsilon^+_a(Q_{1/2}^a+i\bar{Q}_{1/2}^a),\int d^2z
  V_{(0,0)}(\zeta^{AST}_{\mu\nu})\}=\int d^2 z\delta_{\epsilon^+} V+ \epsilon^+_a\zeta^{AST}_{\mu\nu}\int dx e^{\phi/2}
  S^b\partial X^{[\mu} (\gamma^{\nu]})^a_b ,
\end{equation}
where $\delta_{\epsilon^+} V$ is the standard linearized supersymmetry transformation of the \NSNS\
antisymmetric tensor.  In deriving the boundary contribution one
has to carefully take into account the boundary conditions (\ref{bcXPsi}),(\ref{bcSS}) in order to
write all the fields on the boundary in terms of leftmovers.
Hence the supersymmetry transformation of the \NSNS\ AST  produces a nonvanishing
boundary term which is cancelled by the action of the supercharge on
the boundary term (\ref{psipsi}).
\begin{eqnarray}\label{contact2}
  \{\epsilon^+_a(\{Q_{1/2}^a+i\bar{Q}_{1/2}^a), \zeta_{\mu\nu}\int dx
 \psi^\mu\psi^\nu(x)\}&=& \epsilon^+_a\oint dw\; e^{1/2\phi} (\gamma^\rho)^a_b
\partial X_\rho\;\zeta_{\mu\nu}\oint dx \psi^\mu\psi^\nu\nonumber\\
&=& \epsilon^+_a\zeta_{\mu\nu}\oint dx\oint dw {1\over w-x}
  e^{1/2\phi}S^c (\gamma^\rho\gamma^{\mu\nu})^a_c\nonumber\\
&=&  \epsilon^+_a\zeta_{\mu\nu}\int dx
  e^{1/2\phi}(\gamma^{[\mu})^a_b \partial X^{\nu]}S^b+\zeta_{\mu\nu}
  \epsilon^+_a(\gamma^{\mu\nu})^a_b Q_{1/2}^b\nonumber.\\
\end{eqnarray}
The first term in (\ref{contact2}) exactly cancels the boundary term
in (\ref{contact1}). The second term has the form of a shift in the
fermionic collective coordinate.
Here are two comments in order. Firstly the total
derivative in (\ref{contact1}) appears  because the supercharge $Q_{1/2}^a$
in the $s=1/2$ picture was used. It is easy to see that the same
calculation using the supercharge in the $s=-1/2$ picture $Q^{a+}_{-1/2}$
does not lead to a  total derivative term in $\{ Q^{a+}_{-1/2},
V_{(0,0)}(\zeta_{\mu\nu})\}$. This is consistent with  the fact that the
commutator of the supercharge in the $s=-1/2$ picture with the boundary
term (\ref{psipsi})  only produces the
 shift in the fermionic collective coordinate,
\begin{equation}
 \{\epsilon^+_a(\{Q_{-1/2}^a+i\bar{Q}_{-1/2}^a),\int dx
  \zeta_{\mu\nu}\psi^\mu\psi^\nu(x)\}= \zeta_{\mu\nu}
  \epsilon^+_a(\gamma^{\mu\nu})^a_b Q_{-1/2}^b.
\end{equation}
Secondly, the fact that the broken supersymmetry  is nonlinearly
realized on the open string degrees of freedom and acts as a shift in
the fermionic collective coordinate has recently played an important
part in the IIB matrix model \cite{ishibashi}. The supersymmetry algebra for
the zero dimensional reduction of $U(n)$ SYM theory to zero dimensions
\begin{equation}
  \delta A_\mu =\bar{\epsilon}\Gamma_\mu\psi,\quad\delta \psi =
  i[A_\mu,A_\nu]\Gamma^{\mu\nu}\epsilon+\eta. 
\end{equation}
Here $\epsilon$ and $\eta$ corresponds to the unbroken and broken
supersymmetry respectively. Here we suppressed the $U(n)$ Lie algebra
indices but note that the shift $\delta \psi=\eta$ lies in the $U(1)$
part of $U(n)=SU(n)\times U(1)$ which corresponds to the overall
'center of mass' degrees of freedom.  The nonabelian structure $\delta\psi$ becomes manifest
when Chan--Paton factors and open string contact terms are considered
\cite{hamada}. The fact that the supersymmetry is changed by the
presence of boundary terms seems to indicate that the closed string AST
background influences the nonlinearly realized open string supersymmetry.
\section{Picture changing}\label{pic}

In the presence of closed string \NSNS\ antisymmetric
tensor states the  independence of on-shell scattering amplitudes from
the choice of picture makes the presence of 
boundary contact terms (\ref{psipsi}) necessary. In this section the
half plane is conformally mapped into the semi-infinite cylinder and
the D-instanton boundary conditions are encoded in a Boundary state
$\mid B\rangle$ \cite{callanb,lia,greengutperlec}. 
Using the operator formalism the $n$-point amplitude for n \NSNS\
massless tensor states is given in the $(F_2,F_2)$-picture
\begin{equation}\label{Amplitude_n}
A_n=_{F_2,F_2}\langle \zeta_1,k_1\mid V(\zeta_2,k_2)\Delta\cdots \Delta 
V(\zeta_n,k_n)\Delta \mid B\rangle.
\end{equation}
Where the state in  $F_2,F_2$ picture is defined by
\begin{equation}\label{F2F2}
_{F_2,F_2}\langle \zeta_1,k_1\mid
=\zeta^{(1)}_{\mu\nu}  \langle k_1\mid b_{1/2}^\mu
\bar{b}_{1/2}^\nu.
\end{equation}
The closed string propagator is given by
\begin{equation}
 \Delta=
P_{L_0-\bar{L_0}}\frac{1}{L_0+\bar{L}_0-1} \quad,\quad 
P_{L_0-\bar{L}_0}={1\over 2\pi}\int_0^{2\pi} d\sigma
e^{i\sigma(L_0-\bar{L}_0)}.
\end{equation}
To prove the independence of the amplitude of the picture chosen, we
turn the state (\ref{F2F2}) into a state in the picture $(F_1,F_2)$
\begin{equation}\label{stateF2F2}
_{F_2,F_2}\langle \zeta_1,k_1\mid=_{F_1,F_2}\langle
\zeta_1,k_1\mid  G_{-1/2}.
\end{equation}
We want to show that $A_n$ in the $(F_2,F_2)$ is equivalent to $A_n$
in the $(F_1,F_1)$ picture. Inserting (\ref{stateF2F2}) into (\ref{Amplitude_n}), $G_{-1/2}$ is
then moved to the right until it hits the boundary state $\mid
B\rangle$ where it is reflected into a $i\bar{G}_{1/2}$ it is then
moved to the left, where it turns the $(F_1,F_2)$ into a $(F_1,F_1)$
state by
\begin{equation}\label{picF1F1}
_{F_1,F_2}\langle
\zeta_1,k_1\mid  \bar{G}_{1/2}=_{F_1,F_1}\langle \zeta_1,k_1\mid.
\end{equation}
Moving the $G_{1/2}$ to the right has two effects: firstly using
$[L_0,G_r]=-rG_r$ the propagator gets shifted by
\begin{equation}
G_{-1/2}\Delta=\Delta^\prime G_{-1/2}\quad,\quad
\Delta^\prime=P_{L_0-\bar{L}_0-1/2}{1\over L_0+\bar{L}_0-3/2}.
\end{equation}
Secondly we have to show that the commutator $[G_{-1/2},V(m)]$ vanishes
for $m=2,..,n$. Note that the vertex
operators are products of left and right-moving pieces,
i.e. $V=\zeta_{\mu\nu}  V_L^\mu V_R^\nu$. It is well known
that for the physical vertex operator $V$ there exists a
$W=\zeta_{\mu\nu} W_L^\mu V_R^\nu$ and
$W^\prime=\zeta_{\mu\nu} V_L^\mu W_R^\nu$ such that 
$[G_{-1/2},V]=[L_{-1},W]$ and
$[\bar{G}_{1/2},V]=[\bar{L}_{1},W^\prime]$, (see for example \cite{gsw}). The left and right-moving
parts of $W$ are operators of left and right-moving conformal dimension
$1/2$ respectively. Hence the commutator can be replaced by
\begin{eqnarray}\label{cancelled}
[G_{-1/2},V]&\to& (L_{-1}-L_0+1)W-W(L_{-1}-L_0+1/2)\nonumber\\
&+& (L_0-1)W-W(L_0-1/2).
\end{eqnarray}
Where we added and subtracted the second line.  The first line vanishes
because the left-moving part $W_L^\mu$ has conformal dimension of $J=1/2$.
The second line does not contribute to the on-shell scattering amplitude because of the cancelled propagator
argument. Using the level matching projectors  the propagators $\Delta$ and $\Delta^\prime$ can be written
as purely left-moving operators 
\begin{equation}\label{prop}
\Delta={1\over 2(L_0-1/2)}\quad ,\quad \Delta^\prime={1\over 2(L_0-1)}.
\end{equation}
Where in the second line of (\ref{cancelled}) $L_0-1$  cancels the shifted propagator $\Delta^\prime$
to the left and $L_0-1/2$ cancels $\Delta$ to the right. 
When the $G_{-1/2}$ hits the boundary state it is reflected according
to
\begin{equation}
G_{-1/2}\mid B\rangle =i\bar{G}_{1/2}\mid B\rangle.
\end{equation}
The reflected $\bar{G}_{1/2}$ is moved to the left where again shifts
propagators according to 
\begin{equation}
\Delta^\prime \bar{G}_{1/2}=\bar{G}_{1/2}\Delta^{\prime\prime}\quad,\quad
\Delta^{\prime\prime}=P_{L_0-\bar{L}_0}{1\over L_0+\bar{L}_0-2}.
\end{equation}
The commutator $[\bar{G}_{1/2},V]=[\bar{L}_1,W^\prime]$ can be
replaced by
\begin{eqnarray}\label{cancelledbar}
[\bar{G}_{1/2},V]&=& (\bar{L}_{1}-\bar{L}_0)W^\prime-W^\prime(\bar{L}_{1}-\bar{L}_0+1/2)\nonumber\\
&+& (\bar{L}_0-{1\over 2})W^\prime-W^\prime(\bar{L_0}-1).
\end{eqnarray}
Where again we added and subtracted zero. The first line vanishes now
because the right-moving part of $W^\prime$ has $\bar{J}=1/2$. The
second line cancels the propagators to the left and right
\begin{equation}\label{propbar}
\Delta^\prime={1\over 2(\bar{L}_0-1/2)}\quad ,\quad \Delta^{\prime\prime}={1\over 2(\bar{L}_0-1)}.
\end{equation}
Note that $\Delta^\prime$ is different form (\ref{prop}) because the propagators
are now expressed in right-moving oscillators.
When the $\bar{G}_{1/2}$ hits the first state it changes its picture
  (\ref{picF1F1}) and The amplitude $A_n$ is equal to the picture
  changed version where $_{F_2F_2}\langle 1\mid$ is replaced by
  $_{F_1F_1}\langle 1\mid$ and $\Delta$ by $\Delta^{\prime\prime}$. 
This argument fails to be valid because the cancelled propagator for
the vertex closest to the boundary produces a contact term which was
not considered.
\begin{equation}
\cdots\left([G_{-1/2},V_n]\Delta-i[\bar{G}_{1/2},V_n]\Delta^{\prime\prime}\right)\mid
B\rangle
\end{equation}
The explicit form of the vertices for the massless states is given by $V^\mu=(\partial
X^\mu+ik_\rho\psi^\rho\psi^\mu)e^{ikX}$ and $W^\mu=\psi^\mu e^{ikX}$.
Using (\ref{cancelled}) and (\ref{cancelledbar}) the explicit form of
the contact term for massless \NSNS tensor states is given by
\begin{equation}
{1\over 2}\zeta^{(n)}_{\mu\nu}\sum_m\left(
-b^\mu_{m-1/2}(\bar{a}_{m}^\nu+ik^\rho\sum_s\bar{b}^\rho_{m-s}\bar{b}^\nu_{s})+i
\bar{b}^\nu_{m+1/2}({a}_{m}^\mu+ik^\rho\sum_s{b}_{m-s}b^\mu_{s})\right)
\mid
B\rangle.
\end{equation}
The boundary conditions on the modes can be used to write the contact
term entirely in terms of left-moving oscillators. It follows that the
terms trilinear in the fermionic modes vanish and that  the contact term is given by
\begin{equation}\label{vertexvar}
{1\over 2}\zeta^{(n)}_{\mu\nu}\sum_m\left(-{a}_m^\nu
b_{-m-1/2}^\mu+ a_{-m}^\mu{b}_{m-1/2}^\nu\right)\mid B\rangle.
\end{equation}
Note that only an antisymmetric part of $\zeta^{(n)}_{\mu\nu}$
contributes for the boundary term. 
This contact term can be cancelled by an additional boundary term in
$A_n$ given by
\begin{equation}
A_n\to A_n+_{F_2,F_2}\langle 1\mid V(2)\Delta\cdots \Delta 
V(n-1)\Delta
K(n) \mid B\rangle.
\end{equation}
Where the boundary operator is given by
\begin{equation}
K(n)=-i\zeta^{(n)}_{\mu\nu}\sum_{r>0}b^\mu_{-r}\bar{b}^\nu_{-r}.
\end{equation}
The additional term produces a commutator in the picture changing
procedure
\begin{eqnarray}\label{boundvar}
[G_{-1/2}-i\bar{G}_{1/2},K(n)]\mid B\rangle&=&i\zeta^{(n)}_{\mu\nu}\sum_{m<1/2}\left({a}_m^\mu
\bar{b}_{m-1/2}^\nu+i\bar{a}_m^\nu{b}_{m+1/2}^\mu\right)\mid
B\rangle\nonumber\\
&=&{1\over 2}\zeta^{(n)}_{\mu\nu}\sum_{m=-\infty}^\infty\left(- a_m^\mu
b_{-m+1/2}^\nu+a_{m}^\nu b_{-m+1/2}^\mu\right).
\end{eqnarray}
The second line is written in terms of left-moving
oscillators and the factor $1/2$ is introduced in order to double the
range of summation. It is now easy to see that the boundary term (\ref{boundvar})
 cancels (\ref{vertexvar}). This is not the full story since the
cancelled propagator argument for the extra term now produces a
boundary contribution for the vertex $V_{n-1}$
\begin{equation}
\cdots\left([G_{-1/2},V(n-1)]\Delta-i[\bar{G}_{1/2},V(n-1)]\Delta^{\prime\prime}\right)K(n)\mid
B\rangle,
\end{equation}
which can be cancelled
by adding another boundary term $\langle 1\mid
V(1)\cdots V(n-2)K(n-1)K(n)\mid B\rangle$ and so on. 

It is easy to see that all the contact terms are cancelled in the
following sum
\begin{eqnarray}\label{invariantA}
A^\prime_n&=&_{F_2,F_2}\langle 1\mid V(2)\Delta\cdots \Delta 
V(n)\Delta\mid B\rangle\nonumber\\
&+&_{F_2,F_2}\langle 1\mid V(2)\Delta\cdots \Delta 
V(n-1)\Delta
K(n) \mid B\rangle\nonumber\\
&+&_{F_2,F_2}\langle 1\mid V(2)\Delta\cdots \Delta 
V(n-2)\Delta
K(n-1)K(n) \mid B\rangle\nonumber\\
&\vdots&\nonumber\\
&+&
_{F_2,F_2}\langle 1\mid  \Delta
K(2)\cdots K(n) \mid B\rangle +\mbox{other orderings}.
\end{eqnarray}
Where  `other orderings'
denotes the different time orderings on the cylinder which have to be
added. Note that the $K$'s commute since they are made out of creation operators. The corrected amplitude $A^\prime_n$ is now invariant under the picture
changing operation and $A_n$ is equal to (\ref{invariantA}) with $_{F_2,F_2}\langle
1\mid$ replaced by $_{F_1,F_1}\langle 1\mid$ and $\Delta$ replaced by $\Delta^{\prime\prime}$

The appearance of the boundary terms can be understood in the cylinder
frame by considering the inclusion of a boundary term in the action
which is represented as a operator acting on the boundary state
\begin{equation}
Z=\langle 0\mid \exp(iS_{bulk})\exp(iB_{\mu\nu}\oint
  \psi^\mu\bar{\psi}^\nu)\mid B\rangle.
\end{equation}
The scattering amplitude is given by
\begin{equation}
A_n=\prod_{k=1}^n\left(\zeta_{k,\mu_k}\bar{\zeta}_{k\nu_k}{\delta\over
  \delta B_{\mu_k\nu_k}}\right)Z,
\end{equation}
which produces all the terms given in (\ref{invariantA}).

To illustrate the necessity of the contact terms we will consider
amplitude for two \NSNS antisymmetric tensors on the disk with
D-instanton boundary conditions. This amplitude was first calculated
by Klebanov and Thorlacius in \cite{klebanova}.
\begin{equation}
A_2= \int dy \left\langle
 c\bar{c}V_{-2,0}(\zeta^{(1)},k_1)(c+\bar{c})V_{0,0}(\zeta^{(2)},k_2)(iy)\right\rangle={1\over k_1.k_2}{H^{(1)}_{\mu\nu\rho}H^{(2)\mu\nu\rho}}.\label{HH-00}
\end{equation}
Where $H^{(i)}_{\mu\nu\rho}=ik^{(i)}_{[\mu}\zeta^{(i)}_{\nu\rho]}$ for
$i=1,2$. Note
that the important part of the vertex $V_{-2,0}$ in (\ref{HH-00}) is
  up to a super-ghost factor of the same form as $V_{0,0}$ defined in (\ref{vertexNSNS}).
We focus on the  contribution to (\ref{HH-00}) is given by $1/2(\zeta^{(1)}_{\mu\nu}
\zeta^{(2)\mu\nu}-\zeta^{(1)}_{\nu\mu} \zeta^{(2)\mu\nu})$. The two point
function can be calculated in
a different picture defined by
\begin{equation}
  A^\prime_2= \int dy \left\langle
 c\bar{c}V_{-1,-1}(\zeta^{(1)},k_1)(c+\bar{c})V_{0,0}(\zeta^{(2)},k_2)(iy)\right\rangle.
\end{equation} 
Where the vertex operator in the (-1,-1) picture is given by
\begin{equation}
 V_{-1,-1}(\zeta)=\zeta_{\mu\nu} e^{-\phi}e^{-\bar{\phi}} \psi^\mu\bar{\psi}^\nu 
 \end{equation}
 Without the boundary
term the contractions which which produce  $\zeta^{(1)}_{\mu\nu}
\zeta^{(2)\mu\nu}-\zeta^{(1)}_{\nu\mu} \zeta^{(2)\mu\nu}$ are given by
\begin{eqnarray}
A^\prime_2&=&  \zeta^{(1)}_{\mu\nu}
\zeta^{(2)\mu\nu}\int dy
(1-y)^{2k_1.k_2-1}(1+y)^{2k_1.k_2+1}y^{-1-k_2^2}k_2^2\nonumber\\
&&-\zeta^{(1)}_{\nu\mu} \zeta^{(2)\mu\nu}\int dy
(1-y)^{2k_1.k_2+1}(1+y)^{2k_1.k_2-1}y^{-1-k_2^2}k_2^2+\mbox{other terms}.\label{A2-1-1}
\end{eqnarray}
Since these terms are proportional to $k_2^2$ they vanish because of
the on-shell condition for the massless antisymmetric tensor
state $k_2^2=0$. However if one allows the state to be slightly off shell, we
can use the following formula
\begin{equation}
  \int dy
  (1-y)^a(1+y)^by^c=2^{-(2c+2)}{\Gamma({a+1\over2})\Gamma(c+1)\over \Gamma({a+1\over2}+c+1)},
\end{equation}
which is valid for $a+b+2c+2=0$ to find that the integral
(\ref{A2-1-1}) is given by
\begin{equation}
  \zeta^{(1)}_{\mu\nu}
\zeta^{(2)\mu\nu}k_2^2 {\Gamma(k_1.k_2)\Gamma(k_2^2)\over \Gamma(k_1.k_2+k_2^2)}     -\zeta^{(1)}_{\nu\mu} \zeta^{(2)\mu\nu}k_2^2 {\Gamma(k_1.k_2+1)\Gamma(k_2^2)\over \Gamma(k_1.k_2+1+k_2^2)}.    
\end{equation}
The factor of $k_2^2$ is cancelled by a pole coming from the $\Gamma$
function,i.e.  $k_2^2\Gamma(k_2^2)=\Gamma(k_2^2+1)\to
1$ as $k_2^2\to 0$. Hence if we allow the vertex operators to be
slightly off-shell the amplitude (\ref{A2-1-1}) contains the term of
the form $\zeta^{(1)}_{\mu\nu}
\zeta^{(2)\mu\nu}-\zeta^{(1)}_{\nu\mu} \zeta^{(2)\mu\nu}$. But conformal
invariance requires the vertex operators to be on shell and this
continuation can only be regarded to be a trick to get the correct  gauge
invariant result.
Indeed the amplitudes written as functional integrals on the half
plane can be represented in the cylinder frame and the two point
functions in the two different pictures correspond to
\begin{eqnarray}
  A_2&=&_{F_1F_1}\langle \zeta^{(1)}\mid V(\zeta^{(2)})\Delta\mid B\rangle\\
A^\prime_2&=&_{F_2F_2}\langle \zeta^{(1)}\mid V(\zeta^{(2)})\Delta\mid B\rangle.
\end{eqnarray}
 It was argued above that it is necessary to introduce a
boundary term when \NSNS antisymmetric tensor vertex operators are
present. For the amplitude $A_2^\prime$ such a boundary term is given by
\begin{eqnarray}
_{F_2F_2}\langle \zeta^{(1)}_{\mu\nu}\mid  K(\zeta^{(2)})\mid B\rangle&=&\zeta^{(1)}_{\mu\nu}\zeta^{(2)}_{\rho\lambda}\langle
k_1\mid 
b_{1/2}^\mu\bar{b}_{1/2}^\nu(b_{1/2}^\lambda
\bar{b}_{1/2}^\rho- b_{1/2}^\rho \bar{b}_{1/2}^\lambda
) \mid B\rangle\nonumber\\
&=& \zeta^{(1)}_{\mu\nu}
\zeta^{(2)\mu\nu}-\zeta^{(1)}_{\nu\mu} \zeta^{(2)\mu\nu}.
\end{eqnarray}
The inclusion of the boundary term  completes the
gauge invariant result (\ref{HH-00}) which was calculated in a different picture
hence showing that the calculation of the two point function for
\NSNS\ antisymmetric tensor states  in the two pictures  are
equivalent once the boundary terms are included.
\section{Conclusions}\label{conclusions}
In this paper it was  shown that the presence of D-instantons makes
the inclusion of boundary contact terms for \NSNS\ antisymmetric
tensor fields necessary. This  situation  is analogous  to the one
discussed in \cite{greenseiberg}. We have seen that  the contact terms are
necessary to restore gauge invariance and other symmetries. The 
standard  cancelled propagator argument fails because
of kinematic restrictions imposed by  D-instanton boundary conditions.
As indicated in (\ref{opeclosed})  the operator product of a \NSNS\
vertex operator coming close to the boundary produces a total
derivative proportional to a `$0/0$'
expression because of the on-shell condition on the momentum of this
vertex. The boundary contact term removes this term and restores the
gauge invariance. This  term is also necessary for spacetime
supersymmetry and  picture changing symmetry.
The boundary term (\ref{psipsi}) effectively acts as a Lorentz
rotation on spacetime spinors and modifies the action of the closed
string supersymmetries on the open string fields. This fact might have
some interesting consequences when the instanton induced effective
vertices  given by integration over the fermionic collective coordinates
of a D-instanton \cite{greengutperlee} are considered in a nontrivial closed string
background. It would be interesting to investigate wether there is a
connection  with recent suggestions of the
appearance  of central terms in the supersymmetry algebra associated
with p-branes in the worldsheet R-NS formalism using spacetime supersymmetry
charges in different pictures \cite{poyakov} or the incorporation of extra dimensions
using \RR\ matter ghost fields \cite{berkov}.
 
 Note that for D-brane boundary conditions with $p>-1$
the conventional cancelled propagator arguments apply and as stressed
in \cite{greenseiberg} give the same result as the inclusion of
contact terms. The boundary conditions on the open string for a
D p-brane with worldvolume in the $0,1,\cdots, p$ direction is given by

\begin{equation}
  \left(\partial X^\mu+M^\mu_\nu
  \bar{\partial}X^\nu\right)|_{Im(z)=0}=0,\quad \left(\psi^\mu+M^\mu_\nu \bar{\psi}^\nu\right)|_{Im(z)=0}=0
\end{equation}
Where $M^\mu_\nu=diag(-{\bf 1}_{p+1},{\bf 1}_{9-p})$. The OPE of a
closed string AST vertex operator which approaches the D-p-brane
boundary is now given by 
\begin{eqnarray}
  lim_{y\to 0}V(\zeta^{AST}_{\mu\nu})&\sim& \zeta^{AST}_{\mu\nu}M^{\nu}_{\rho}{1\over
  (y)^{k.M.k+1}}k.M.k \psi^\mu\psi^\rho (x)+o(y^0)\nonumber \\
&=& \left({\partial\over \partial y} y^{k.M.k}\right)
  \zeta^{AST}_{\mu\nu}M^\nu_\rho \psi^\mu\psi^\rho (x)+o(y^0)
\end{eqnarray}
And since $k.M.k= k_0^2-k_1^2+\cdots k_{p}^2-k_{p+1}^2-\cdots k_9^2$
it is possible if $p>-1$ to choose a physical on shell momentum $k^\mu$ such that
$k.M.k>0$ and the boundary contact term vanishes. Hence the contact
terms are important for D-branes only if the closed string momentum
$k^\mu$ is forced to vanish, which might be important when wrapped
euclidian branes are considered. 
\vskip 0.8cm
\noindent{\bf  Acknowledgments}:
\vskip 0.2cm
\noindent I would like to thank M.B. Green  for
useful conversations and suggestions.

\end{document}